%% file: main.tex
\documentclass{article}

\usepackage[dblblindworkshop,final]{neurips_2026}

\usepackage[utf8]{inputenc}
\usepackage[T1]{fontenc}
\usepackage{url}
\usepackage{booktabs}

\usepackage{graphicx}
\usepackage{amsmath}
\usepackage{amssymb}
\usepackage{nicefrac}
\usepackage{microtype}
\usepackage{xcolor}
\usepackage{subcaption}
\usepackage{multirow}

\usepackage{hyperref}

\definecolor{lightsim2grid}{HTML}{774703}
\definecolor{pgm}{HTML}{A96404}
\definecolor{rustpower}{HTML}{DB8206}
\definecolor{exapf}{HTML}{F99C1A}
\definecolor{pandapower}{HTML}{FAB24C}
\definecolor{olf}{HTML}{FCC77E}
\definecolor{powermodels}{HTML}{FDDDAF}
\definecolor{GNS}{HTML}{0C4697}
\definecolor{pfnet}{HTML}{136DEC}
\definecolor{canospf}{HTML}{68A2F3}
\definecolor{gpusim2grid}{HTML}{8A0F24}
\definecolor{exapfGPU}{HTML}{E6193C}

\title{Benchmarking CPU, GPU, and ML Power-Flow Solvers from 14 to 70,000 Buses}
\workshoptitle{AI Foundations for Power Grids}
\author{%
Benjamin Donnot \quad Sami Tazi \quad Eva Boguslawski \quad Nourredine Henka \\
RTE France \\
\texttt{{benjamin.donnot, sami.tazi, eva.boguslawski, nourredine.henka} @rte-france.com}
}

\begin{document}

\maketitle

\begin{abstract}
Learned surrogates for power-flow (PF) computation are usually motivated by speed, compared against a single CPU-based baseline. We show this comparison is fragile: across 7 open-source CPU-based solvers evaluated on identical inputs, throughput varies by roughly 2.5–3 orders of magnitude, so the choice of baseline alone can flip the conclusion. We build a benchmark spanning 9 physics-based solvers – 7 CPU-based and 2 GPU-based, including a new batched GPU solver, gumsim2grid – and 3 published ML surrogates, evaluated under a shared, independently implemented KCL-mismatch check across 15 grids from 14 to 70,000 buses. On the PF task itself, a sufficiently parallelized CPU solver or gpusim2grid matches or exceeds the throughput of the ML surrogates we test while remaining orders of magnitude more accurate; we find no evidence that these GNN-style surrogates are yet competitive for PF specifically, though prior work reports genuine gains on related but more combinatorial tasks such as optimal power flow or power grid control. Beyond this finding, we hope the benchmark suite – code and reference evaluator released at an anonymized repository – helps close a broader reproducibility gap in this space, where batched GPU-based PF solvers are rarely open-sourced or evaluated against a common set of baselines.
\end{abstract}

\section{Introduction}

Power flow (PF) solving is a computational bottleneck both for grid operations and for ML pipelines that rely on PF solvers to generate training data at scale: rising renewable penetration is increasing the number of instances operators must evaluate, and recent large-scale neural surrogate efforts report data-generation budgets that are themselves a substantial compute cost. 

A large and active line of machine-learning work responds to this need for speed with learned surrogate models \citep{rivera2025pfdelta,puech2026genco}. These comparisons are established against CPU-based baselines, since batched GPU-based solving has not, until recently, been practical: the sparse factorizations at the core of a Newton-Raphson solve mapped poorly onto GPU hardware at the scale of a single instance. Batched sparse solvers such as cuDSS \citep{nvidia2026cudss} make GPU-based physics solving newly practical, which raises the question we investigate: how does the ML throughput case, established mainly against CPU baselines, hold up once that option exists?

Comparisons between ML models and a single CPU-based baseline can be misleading, as CPU solvers themselves vary substantially in performance. We observe throughput differences of up to $\sim$2.5–3 orders of magnitude across CPU implementations
, so comparing against just the slowest or fastest baseline can yield opposite conclusions on whether ML is ``needed’’ for speed. We built a benchmark spanning more than 3.5 orders of magnitude in grid size (14 to 70{,}000 buses): 7 CPU-based tools, 2 GPU-based physics solvers (including gpusim2grid, introduced here), and 3 ML surrogates, across multiple languages and academic/industrial tools – to our knowledge the first time PF computation is evaluated at this scale against this many alternatives on common ground. Code, including the reference KCL evaluator (Section~\ref{sec:formulation}) used to check every solver on equal terms, is available at an anonymized repository\footnote{Anonymized code: \url{https://github.com/Grid2op/gpusim2grid}.}.

We find that a sufficiently parallelized CPU-based solver, or gpusim2grid batch physics solver (Section~\ref{sec:suppl_gpusim2grid}), matches or exceeds the throughput of the ML surrogates that we evaluate on the power-flow task while being orders of magnitude more accurate. GENCO \citep{puech2026genco} reports genuine gains for learned surrogates on more computation-heavy tasks such as Optimal Power Flow (OPF) (Section~\ref{sec:related}); we read the two as complementary rather than contradictory. In summary, our two contributions are: (1) the benchmark suite described above, addressing the reproducibility gap left by closed-source GPU accelerators in this space; and (2) gpusim2grid itself, matching the precision of CPU-based solvers and exceeding the throughput of the ML surrogates.

\section{Problem Setting and Related Work}
\label{sec:related}

\subsection{Power flow formulation}
\label{sec:formulation}

We consider the classical AC power-flow problem, whose goal is to find, for each bus $i$, the complex voltage $v_i$ satisfying the equation:
\begin{equation}
v\odot(Y.v)^* = s
\label{eq:powerflow}
\end{equation}
where $Y.v$ is the matrix product of the admittance matrix $Y$ and $v$, $\cdot^*$ is the conjugate transpose, and $\odot$ is the Hadamard product. We follow the bus-type conventions and mismatch definition of \citet{zimmerman2011matpower} (also used by \citet{rivera2025pfdelta,thurner2018pandapower}).



\subsection{Related work}

GPU-based acceleration of power flow is an active line of work. ExaPF.jl \citep{schanen2021exapf} moves Newton-based AC power-flow solves onto GPU.
gpusim2grid instead targets the complementary, batched regime relevant to contingency analysis and training-data generation, alongside three other efforts in this same batched, sparsity-aware GPU regime: SABLE \citep{song2026sable},
a cuDSS-based batched refactorization scheme, not yet open-sourced to our knowledge. A concurrent JAX/GMRES implementation \citep{zhou2026jaxbatched} and an earlier batched sparse-LU approach \citep{wang2021seganbatched} both lack publicly available code, to our knowledge. None release code or are evaluated against a common set of baselines – precisely the reproducibility gap our benchmark suite tries to close.

Concurrently, GENCO \citep{puech2026genco} proposes a neural solver unifying PF, OPF, and State Estimation (ST), reporting up to $30\times$ throughput over a PF baseline in a hardware-matched CPU/GPU comparison, mainly on grids $\geq$2,000 buses at a looser tolerance than our CPU baseline. We do not dispute this comparison; together with our results (Section~\ref{sec:results}), it suggests learned surrogates show their clearest gains on OPF or ST rather than plain PF, where gpusim2grid is, to our knowledge, the only openly released implementation benchmarked across a comparably wide grid size and solver range.


\section{Experiments and results}
\label{sec:results}

\subsection{Common experimental settings}
\label{sec:common_settings}

Every solver benchmarked below solves the same AC power-flow system on the same 15 grids (14 to 70{,}000 buses)\footnote{The full list of grids used is available in Section~\ref{sec:supp_grids}.}: 7 open-source Newton-Raphson CPU-based tools – pandapower (\textbf{pp}) \citep{thurner2018pandapower}, \textbf{powermodels} \citep{coffrin2018powermodels}, exapf-CPU (\textbf{exapf}) \citep{schanen2021exapf}, pypowsybl/OLF (\textbf{olf}) \citep{pypowsybl2026,openloadflow2026}, lightsim2grid (\textbf{ls}) \citep{lightsim2grid}, power grid models (\textbf{pgm}) \citep{Xiang_PowerGridModel_power-grid-model,Xiang2023_powergridmodels}, and \textbf{rustpower} \citep{cheng2026rustpower} – 2 open-source Newton-Raphson GPU-based physics solvers (\textbf{exapf-GPU}, \textbf{gpusim2grid}), and 3 ML surrogates (PowerFlowNet~\citep{lin2024powerflownet}, GNS-S~\citep{donon2020gns}, CANOS-PF~\citep{piloto2024canos}). We used \citet{song2026sable}'s data-sampling methodology by default.

\subsection{Physical solver validation}
\label{sec:baseline_calibration}
\paragraph{Experimental design.} This first set of experiments checks that all benchmarked solvers solve the same problem, to ensure fairness. To that end, we wrote an independent script implementing the KCL evaluator of Section~\ref{sec:formulation} (detailed in Section~\ref{sec:suppl_kcl_checker} and developed directly from the matpower reference manual \citep{zimmerman2011matpower}) to check every solver’s output under a common configuration (per-solver details in Section~\ref{sec:supp_solver_config}).

We report the ``mean/worst-case KCL mismatch’': with $\text{mis} = v\odot(Y.v)^* - s$, the violation vector is $[\Re(\text{mis}(pv)), \Re(\text{mis}(pq)), \Im(\text{mis}(pq))]$; the mean/worst-case mismatch is the average/max, over the dataset, of its mean absolute value over the grid.

We sampled 9{,}000 independent scenarios per grid, identical across baselines, solved at tolerance $10^{-8}$ \citep{song2026sable}; mismatch is computed over converged samples only.

\paragraph{Results.}

At 5 iterations, gpusim2grid is as precise as the reference CPU solver in all cases (9,000 scenarios $\times$ 15 grids): worst-case KCL mismatch is $2.31\times10^{-10}$, below the $10^{-8}$ tolerance (Table~\ref{tab:kcl_9k}); all reported timings in the next sections use 5 iterations for gpusim2grid. All Newton-Raphson-based solvers solve the PF problem, independently validated by our evaluator (full per-solver results in Table~\ref{tab:suppl_kcl9k_fll}, Section~\ref{sec:supp_kcl_baselines}).

\begin{table}[ht]
    \centering
    \footnotesize
\begin{tabular}{lllrr}
\toprule
 Solver       & Grids (max size)   & Convergence rate   &   Mean mismatch &   Worst-case mismatch \\
\midrule
 pandapower (ref) & 15 (70000)         & 100.00\%            &        5.61e-11 &              1e-08    \\
 \midrule
 gpusim2grid (4 iter) & 15 (70000)         & NA                  &        1.05e-09 &              0.000116 \\
 \textbf{gpusim2grid (5 iter)} & \textbf{15 (70000)}         & NA                  &        \textbf{8.91e-14} &              \textbf{2.31e-10} \\
 gpusim2grid (6 iter) & 15 (70000)         & NA                  &        8.8e-14  &              1.04e-10 \\
\bottomrule
\end{tabular}
    \caption{KCL mismatch, pandapower (reference) vs.\ gpusim2grid. 5 iterations is the lowest budget meeting the $10^{-8}$ tolerance used throughout. Full per-solver table in Section~\ref{sec:supp_kcl_baselines}.}
    \label{tab:kcl_9k}
\end{table}



\subsection{Timings on generation and demand varying scenarios}
\label{sec:time_series}

\begin{figure}[ht]
\centering
\includegraphics[width=0.65\linewidth]{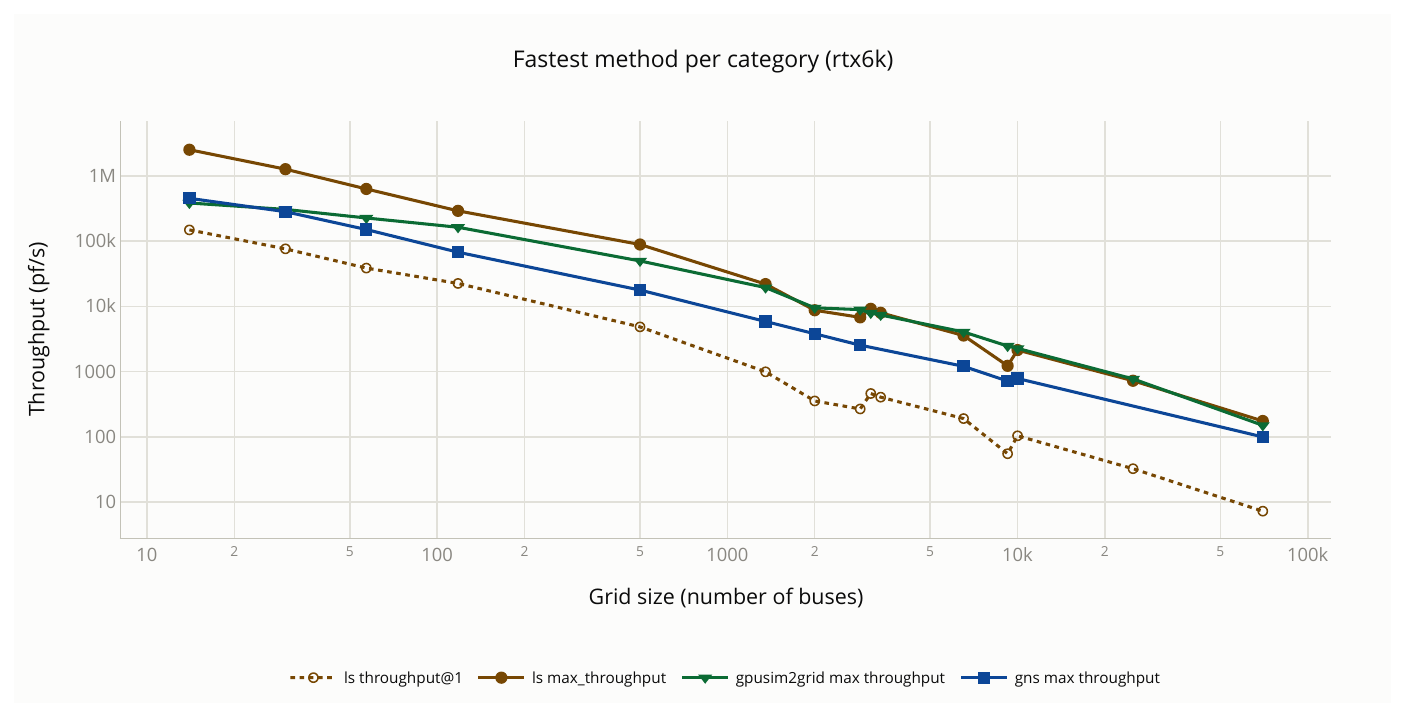}
\caption{Fastest solver per category, rtx6k machine: lightsim2grid (best NR CPU), gpusim2grid (best NR GPU), and GNS (best ML surrogate). gpusim2grid tracks lightsim2grid, and both outperform the ML surrogate across grid sizes. }
\label{fig:exp_best}
\end{figure}

\paragraph{Experimental design.} This section measures each solver’s maximum throughput – Newton-Raphson on CPU, Newton-Raphson on GPU, or a learned surrogate on GPU – pushed to its limit, on a task common in the literature (\textit{e.g.}, \citep{song2026sable}): varying the generation and load profile for a fixed topology (details in Section~\ref{sec:suppl_cpu_spread}). Following \citep{puech2026genco}, we report the best achieved throughput for each of the 12 benchmarked methods\footnote{Methodology and machine specs in Section~\ref{sec:suppl_cpu_spread}.}; Figure~\ref{fig:exp_best} shows the fastest method in each of the 3 categories (NR/CPU, NR/GPU, learned model).

This figure is favorable to the benchmarked learned models: a fixed architecture (the one proposed in the PF-$\Delta$ paper, trained only up to 2000 buses) is reused unmodified on every grid in our suite, without retraining or resizing for the largest cases. Our throughput comparison therefore bounds what this specific, non-adapted surrogate achieves, not necessarily what a resized or re-trained model could reach on grids far outside its training range, in either throughput or accuracy. This part of our benchmark focuses on throughput; accuracy is compared in Section~\ref{sec:pfdelta_accuracy}.



\paragraph{Results.}
Figure~\ref{fig:exp_best} shows a large gap between the fastest CPU solver (lightsim2grid for this benchmarked task) on a single core and at maximum throughput\footnote{Number of cores chosen to effectively maximize the number of power flows per second} – on average 16$\times$ faster – which matters when comparing a GPU to one CPU core rather than to the full CPU: the fastest learned model (the “Graph Neural Solver”) beats 1 CPU core of the fastest CPU baseline, but not the full CPU. The evaluated NR algorithm on the GPU (gpusim2grid), however, matches the CPU’s throughput on the largest grids. 

\subsection{PF-Delta and accuracy}
\label{sec:pfdelta_accuracy}
\paragraph{Experimental design.} Table~\ref{tab:pfdelta-accuracy-mean} reports the Power Balance Loss \citep{rivera2025pfdelta} for pandapower and gpusim2grid on the PF-$\Delta$ datasets\footnote{More details about the data sampling strategy in the original paper here \citep{rivera2025pfdelta}, especially section "4.1 Data Generation"}, alongside published accuracy numbers for the best ML surrogates among PowerFlowNet, Graph Neural Solver, CANOS-PF and GENCO Base (the best one) in their best configuration, from their source publication.

\paragraph{Results.}

\begin{table}[ht]
    \centering
    \footnotesize
\begin{tabular}{lrrrr}
\toprule
 Model & $N$ & $N$-1 & $N$-2 & Close-to-infeasible (C2I) \\
\midrule
 pandapower (ref.)  & 1.1e-6 & 8.5e-7 & 7.5e-7 & 7.2e-3\textsuperscript{*} \\
 gpusim2grid (ours)  & 4.9e-14 &  2.3e-8  & 3.1e-10  & 4.5e-4 \\
GENCO Base\textsuperscript{$\dagger$} & 2.7e-3 $\pm$ 2.1e-3 & 4.5e-3 $\pm$ 3.4e-3 & 5.8e-3 $\pm$ 4.3e-3 & 2.2e-1 $\pm$ 8e-2 \\
\bottomrule
\end{tabular}
    \caption{Power Balance Loss (mean) [p.u.], IEEE-118. ML row: best ML model (GENCO Base, \citet{puech2026genco}) on its best-case task (1.3/2.1), PF-$\Delta$ dataset. pandapower/gpusim2grid: run by us on the same dataset (includes diverging power flows). \textsuperscript{*}pandapower's C2I mean is skewed by three non-converging samples; excluding them gives $1.89\times10^{-6}$. \textsuperscript{$\dagger$}Run by \citet{puech2026genco} on their own hardware, not ours.}
    \label{tab:pfdelta-accuracy-mean}
\end{table}

In this setting – most favorable to the ML surrogates (best model, evaluated on its best-performing task), matching train/test distributions – pandapower and gpusim2grid are far more accurate on case118 across $N$/$N$-1/$N$-2: over $1000\times$ vs.\ the most accurate ML model (GENCO Base, 2.7e-3, against pandapower’s 1.1e-6). C2I results are skewed by three non-converging outliers (see table note); gpusim2grid achieves near-full convergence within the same budget.

\section{Conclusion and discussion}
\label{sec:discussion}
\label{sec:limitations}

Our results show a well-parallelized CPU-based solver or gpusim2grid outperforms the GNN-based ML surrogates evaluated for power flow specifically. This does not concern all learned methods (or architectures), and our findings are consistent with GENCO’s reporting higher gains on OPF and state estimation \citep{puech2026genco}. These results do not concern the use of GNNs for other tasks such as tertiary voltage control \citep{donon2026tvc} or grid topology optimization \citep{hassouna2025graphrl}.

All the learned methods benchmarked are GNN-based, and other architectures, e.g.\ transformers, might trade off speed and accuracy differently. We also report a single best-observed throughput per solver/grid rather than repeated-run statistics, following common practice at this scale (e.g.\ \citet{puech2026genco}, \citet{song2026sable}); repeating our 630M solves multiple times each was not tractable within our compute budget. Our benchmark was also limited by the machines we had available, 
(\textit{e.g.}, a more recent or larger CPU or GPU). The synthetic data used here also does not fully capture real-grid complexity; evaluating on GENCO’s realistic dataset is an interesting future direction.

\paragraph{LLM usage disclosure.}
This manuscript’s text was drafted and revised with the assistance of a
large language model (Claude, Anthropic), which also helped write the
data-analysis and figure-generation scripts used to
visualize the benchmark results. Beyond manuscript preparation, the
model was also used as part of the research method itself: it assisted
the authors in implementing the gpusim2grid solver
(Section~\ref{sec:suppl_gpusim2grid}) – the batching strategy and its
CUDA/cuDSS-based implementation – under the authors’ direction,
review, and testing at every stage. It was not involved in the
benchmark’s experimental design (choice of solvers, grids, sampling
protocol, or metrics) nor in producing any reported numerical result:
all timing runs were launched by the authors, accuracy figures were generated by the authors’ own
pipeline and independently verified against the reference KCL
evaluator of Section~\ref{sec:formulation}. All LLM-assisted content,
including code, was reviewed by the authors, who take full
responsibility for the manuscript’s accuracy and integrity.

{\small
\bibliographystyle{plainnat}
\bibliography{references}
}

\newpage
\input{domain_checklist.tex}

\section{Supplementary material}

\subsection{Power grids used for this experiment}
\label{sec:supp_grids}
We tried to use available grids in the literature that were used in published benchmarks, for example in \citep{song2026sable} or \citep{puech2026genco}, while also using the smallest and largest possible grids. We finally used:

\begin{itemize}
    \item case14, case\_ieee30, case57, case118 from \citep{zimmerman2011matpower} have also been used by the PF-$\Delta$ paper
    \item case1354pegase, case9241pegase from \citep{fliscounakis2013contingency}, case3120sp from \citep{zimmerman2011matpower} and case\_ACTIVSg10k from \citep{birchfield2017grid} were used in SABLE
    \item case\_ACTIVSg500, case\_ACTIVSg2000, case\_ACTIVSg25k, case\_ACTIVSg70k from \citep{birchfield2017grid}, case2869pegase, case6515rte from \citep{fliscounakis2013contingency} and case3375wp from \citep{zimmerman2011matpower}
\end{itemize}

They cover a huge range of sizes and complexity, spanning from the ``toy model'' case14 up to the ACTIVSg70k test case, built to represent the eastern United States (52 areas in total) with 70 thousand buses.

\subsection{Independent KCL evaluation metrics}
\label{sec:suppl_kcl_checker}
To make sure that we were comparing solvers solving the exact same problem, we built a dedicated script that tests whether or not a given solution (provided by any solver) meets Kirchhoff's Current Law (KCL).

This validation takes as input a matpower case file (.m file), which is the file most of the benchmarked solvers can read.

It then builds the $Y$ matrix from the bus (columns GS and BS) (adding the shunt diagonal entries) and the branch matrix (both diagonal and non-diagonal entries) using the BR\_B, BR\_X, BR\_R, TAP and SHIFT columns (only for lines that are actually connected BR\_STATUS $\neq 0$), as described in the \href{https://matpower.org/docs/MATPOWER-manual-7.0b1.pdf}{MATPOWER manual}, Section ``3 Modeling", pages 22 and following, especially subsection ``3.2 Branches".

The $s$ vector is built similarly from the same manual, especially subsection ``3.3 Generators'' and ``3.4 Loads'' (pages 25 and following).

Finally, the type of each bus (``slack'', ``pv'' or ``pq") is not read directly from the matpower bus data, because that is not what most of the tested CPU baselines do today. To determine the type of a bus (``slack'', ``pv'' or ``pq"), the following rule is used:

\begin{itemize}
    \item if it is tagged as ``slack'' in the matpower bus data, then it is the slack; we do not change that
    \item a bus is said to be ``pv'' if it has a \textbf{connected} generator (GEN\_STATUS $\neq 0$), discarding the input matpower data for this field
    \item if a bus is neither the ``slack'' nor ``pv'', it is declared ``pq''.
\end{itemize}

\subsection{Dive into gpusim2grid}
\label{sec:suppl_gpusim2grid}

gpusim2grid is built on top of CUDA and cuDSS \citep{nvidia2026cudss}, NVIDIA's batched sparse direct-solver library, chosen for being, to our knowledge, the most mature batched sparse direct-solver library currently available for GPU. The trade-off is that cuDSS is CUDA-only, so gpusim2grid as implemented does not run on AMD ROCm hardware, unlike exapf-GPU in our benchmark; targeting a ROCm-compatible batched sparse solver, should one reach comparable maturity, is straightforward in principle but out of scope here.

In particular the gpusim2grid model implements a full AC powerflow on the GPU. A converged grid (on the CPU) is uploaded once to the GPU, which then:
\begin{itemize}
    \item extracts from this converged grid all the elements required to solve the system \ref{eq:powerflow}: $Y$, $s$, ``slack'', ``pv'' and ``pq'' indices
    \item then extracts the data required to build the ``Jacobian'' $J$ of the system\textsuperscript{*} and represent it in cuDSS
    \item stacks multiple versions (exactly batch-size many) of $Y$, $J$ and $v$, giving $\mathbf{Y}$, $\mathbf{J}$ and $\mathbf{v}$. This can be done in an ``embarrassingly parallel'' manner because only the values of $Y$ and $J$ need to be stacked; the sparsity pattern is the same across the whole batch.
\end{itemize}

\textsuperscript{*} The complete details on how a Newton-Raphson-based powerflow works are out of scope for this paper. What matters for this work is that solving an AC powerflow with the Newton-Raphson algorithm requires solving a different linear system at each iteration, in the form $J.dx = \text{mis}$, where mis is the ``KCL violation vector'' defined in Section~\ref{sec:formulation} and $J$ is a sparse matrix whose sparsity pattern is derived from $Y$, ``pv", ``pq'' and ``slack'', and whose values can be computed by a closed-form formula given $Y$, $s$, and the complex voltage vector at the given iteration.

Then the ``solve'' part will take place, all on the GPU, without any CPU-GPU transfer during computation. For a given number of iterations (user specified, uniform across the elements in the batch):

\begin{enumerate}
    \item the values of the stacked $\mathbf{J}$ are updated from $\mathbf{v}$ and $\mathbf{Y}$ for all systems in the batch, as if it were a single (block-diagonal) matrix.
    \item cuDSS then solves the stacked linear system $\mathbf{J}.dx = \mathbf{mis}$ for the whole batch at once
    \item the data for each individual system in the batch is read back, which is used to build the updated (stacked) $\mathbf{v}$ complex vector at each bus
    \item the KCL mismatch is computed for all systems in the batch, giving the vector $\mathbf{mis}$ used to restart at step 1.
\end{enumerate}

This way of solving the powerflow with the Newton-Raphson algorithm is beneficial (in terms of computation time) if enough data is used at step 2 above (the most time-consuming step); otherwise, the setup cost might decrease throughput. This can explain why, in Figure~\ref{fig:exp_best}, the gpusim2grid method does not perform as well as other cpu baselines on the 4 or 5 smallest grids: the GPU might not be saturated in these cases.

With this formulation, this model can already handle, in a single batch:

\begin{itemize}
    \item varying active generation and demand values
    \item varying generation voltage setpoint
    \item varying reactive demand
    \item disconnection of an arbitrary number of branches (known as $n-k$), including power line or transformer disconnections that would island part of the grid (in this case the islanded part is not solved)
\end{itemize}

This is why, for the experiments of Section \ref{sec:pfdelta_accuracy} we could not run the whole dataset in one pass. For now there is no possibility to batch generator disconnection directly (as it would change the ``pv'' to ``pq'' attribution of some buses). We had two solutions to handle this:

\begin{itemize}
    \item Either we kept ``everything in one batch'' approach, and in this case we had to slightly modify the input data to model the transformer connecting the generator to the bus. This could be done in a way to make the modified model and the new model identical, ensuring the KCL evaluator would still report full convergence for gpusim2grid and then use the gpusim2grid possibility to disconnect these transformers (that would then also disconnect the generator behind them) in the batch.
    \item or we could more simply run multiple instances of the gpusim2grid simulator, each initialized from the state where the generator status are fixed. We then run all the data in "one batch" for every generator disconnection pattern in the test sets. This simpler approach has been preferred for this paper, though it makes the gpusim2grid much slower in this setting: each instance of gpusim2grid needs to copy $Y$, $s$, build the jacobian etc. which (for small grid like case118) takes a lot of time when compared to the actual powerflow solves.
\end{itemize}

What could be possible (but is not yet implemented) is to also include, for a given batch, if known before computation, more profound topological modifications (bus splitting/merging, moving an element to a different bus) and changes to bus status (\textit{e.g.}, ``pv'' to ``pq'' or conversely). This could be done without changing the core of the architecture and will be part of future developments.

An ability that might also interest current state-of-the-art data-processing pipelines \textit{e.g.} in \citep{puech2026genco} is to slightly change the physical parameters of the power lines. This has not been done yet but could be addressed with, in some cases, parameter sampling directly on the GPU for an even higher throughput.

What is not yet addressed is the capacity to batch data from multiple grids of different sizes in the same call, or to modify parameters mid-solve (\textit{e.g.}, disconnecting a line or a generator between iterations 2 and 3).

\subsection{Physical solvers configuration used}
\label{sec:supp_solver_config}
Having all 7 solvers solve exactly the same AC power flow problem has not been an easy task.
We had to tweak the default configuration of some of them to make sure every element was represented exactly as in Equation~\ref{eq:powerflow}.

We think that finding these exact parameters is one contribution of this paper. All the code for the evaluation of all models is provided on GitHub\footnote{Anonymized version available here:  \url{https://github.com/Grid2op/gpusim2grid/blob/main/20260831_neurips_paper/scripts/kcl_checker.py}}. We encourage anyone interested in the details to check there for the actual implementation. We only expose here a summary of what has been done:

\begin{itemize}
    \item \textbf{pandapower}: has been one of the most difficult to use because it uses, by default, a different representation for transformers and changes the order of the elements (the $k$\textsuperscript{th} bus in the matpower case file is not necessarily the $k$\textsuperscript{th} in pandapower). To solve the first issue, we model transformers explicitly as an ``impedance'' pandapower element (rather than as transformers, as done by default). We also wrote a script to ``align'' the IDs of the buses, loads, and generators with the matpower data.
    \item \textbf{powermodels}: actually one of the easiest to use. It is directly ``aligned'' with matpower data by default. We compute powerflows here with the \texttt{compute\_ac\_pf} function (rather than the costlier \texttt{solve\_ac\_pf}) for grids of all sizes (the GENCO paper instead used Ipopt, a costlier solver, for the largest grids, 2k and 10k buses). For these largest grids, like \citet{puech2026genco}, we noticed some ``unrealistic'' solutions, though still marked as valid by PowerModels. We still report computation time for these cases, since they still meet the KCL.
    \item \textbf{exapf}: also easy to use; the default importer matches the ``matpower equations'' bit for bit.
    \item \textbf{pypowsybl}: this CPU solver only imports the matpower ``.mat'' format (not .m), so we used a converter (via a local matpower installation under the free Octave language) to convert the data. It also required changing the order of the elements (especially buses and generators) by relying on the imported names. We also had to use dedicated powerflow parameters to make pypowsybl run the exact powerflow at the given tolerance: by default, it solves a more complex problem that accounts for generator reactive limits or slack distribution, for example. These more realistic features are out of scope for this benchmark and were switched off.
    \item \textbf{lightsim2grid}: also among the easiest. Matpower data is read natively and the element order is preserved.
    \item \textbf{power grid models} (called ``pgm'' in the scripts, to distinguish it from the similarly named ``powermodels") was also quite challenging. It was primarily developed for distribution grids (DSO), whereas matpower data is primarily transmission-focused (TSO). We wrote a full converter from matpower data to power grid models, partly relying on the pandapower converter (after the pandapower fixes above) and forcing some parameters to specific values (e.g., for the slack, or disabling generator reactive limits). Being designed mainly for the smaller grids typical of DSOs, this model does not converge for the largest grids ($\geq 2000$ buses); we plan to raise this on the Power Grid Model GitHub.
    \item \textbf{rustpower}: has no direct interface for reading the matpower format, but accepts lower-level inputs directly ($Y$, $s$, ``pv", ``pq"), which we extracted from lightsim2grid as the best-suited source. We also found (and verified in the rustpower code) that we needed to pass the transpose of the $Y$ matrix for results to match matpower's\footnote{$Y$ matrices are not symmetric in the presence of phase-shifting transformers, i.e., when the SHIFT column is nonzero in the original matpower data.}. We intend to open an issue on their GitHub for this finding.
\end{itemize}

\subsection{Solver names and conventions}
\label{sec:supp_solver_names}
When trying to benchmark all these solvers, we faced an issue: how to name all the different solvers. Finding a good name for ``powermodels'' and ``power grid models", or ``pandapower'' (often abbreviated ``pp") and ``pypowsybl'' (also abbreviated ``pp"), can be confusing. We therefore adopted the following convention:

\begin{itemize}
    \item \textcolor{pandapower}{pp}: pandapower
    \item \textcolor{powermodels}{powermodels}: powermodels
    \item \textcolor{exapf}{exapf}: exapf, in CPU mode only
    \item \textcolor{olf}{olf}: pypowsybl, with its default ``open load flow'' (abbreviated OLF) powerflow solver
    \item \textcolor{lightsim2grid}{ls}: lightsim2grid
    \item \textcolor{pgm}{pgm}: power grid models, its standard abbreviation
    \item \textcolor{rustpower}{rustpower}: rustpower
    \item \textcolor{exapfGPU}{exapf-gpu}: exapf, in GPU mode only
    \item \textcolor{gpusim2grid}{gpusim2grid}: the proposed Newton-Raphson-based model running on GPU
    \item \textcolor{GNS}{gns}: the ``graph neural solver'' model \citep{donon2020gns}, with implementation and dimensions taken from the PF-$\Delta$ codebase
    \item \textcolor{pfnet}{pf-net}: the ``powerflownet'' model \citep{lin2024powerflownet}, with implementation and dimensions taken from the PF-$\Delta$ codebase
    \item \textcolor{canospf}{canos}: the ``canos-pf'' model~\citep{piloto2024canos}, with implementation and dimensions taken from the PF-$\Delta$ codebase
\end{itemize}

The coloring of the graphs also has a meaning:
\begin{itemize}
    \item \textcolor{lightsim2grid}{brown} to \textcolor{powermodels}{yellow} color corresponds to Newton-Raphson-based models running on CPU
    \item \textcolor{gpusim2grid}{burgundy} to \textcolor{exapfGPU}{red} corresponds to Newton-Raphson-based models running on the GPU
    \item \textcolor{GNS}{dark blue} to \textcolor{canospf}{light blue} corresponds to machine learning-based models
\end{itemize}

\subsection{Full results of the KCL mismatch for each tested tool}
\label{sec:supp_kcl_baselines}

\begin{table}[]
    \centering
\begin{tabular}{lllrr}
\toprule
 Solver       & \#Grids (max size)   & Cvg. rate   &   Mean mismatch &   Worst-case mismatch \\
\midrule
 pp           & 15 (70000)         & 100.00\%            &        5.61e-11 &              1e-08    \\
 powermodels  & 14 (25000)         & 78.57\%             &        9.78e-13 &              1.26e-08 \\
 exapf        & 14 (25000)         & 100.00\%            &        9.11e-14 &              8.27e-11 \\
 olf          & 15 (70000)         & 100.00\%            &        4.4e-10  &              1.75e-06 \\
 ls           & 15 (70000)         & 100.00\%            &        1.4e-13  &              1e-10    \\
 pgm          & 5 (500)            & 99.88\%             &        1e-14    &              1.11e-12 \\
 rustpower    & 15 (70000)         & 100.00\%            &        1.31e-10 &              8.65e-09 \\
 \midrule
 gpusim2grid (0 iter) & 15 (70000)         & NA                  &        0.0142   &              4.39     \\
 gpusim2grid (1 iter) & 15 (70000)         & NA                  &        0.00815  &             45.4      \\
 gpusim2grid (2 iter) & 15 (70000)         & NA                  &        0.000365 &              4.5      \\
 gpusim2grid (3 iter) & 15 (70000)         & NA                  &        3.08e-06 &              0.0806   \\
 gpusim2grid (4 iter) & 15 (70000)         & NA                  &        1.05e-09 &              0.000116 \\
 gpusim2grid (5 iter) & 15 (70000)         & NA                  &        8.91e-14 &              2.31e-10 \\
 gpusim2grid (6 iter) & 15 (70000)         & NA                  &        8.8e-14  &              1.04e-10 \\
 gpusim2grid (7 iter) & 15 (70000)         & NA                  &        8.8e-14  &              1.04e-10 \\
\bottomrule
\end{tabular}
    \caption{Full results of the KCL experiments for the 7 Newton-Raphson reference baselines and the gpusim2grid model for each of the tested iterations. 9,000 data points were sampled independently.}
    \label{tab:suppl_kcl9k_fll}
\end{table}

Table~\ref{tab:suppl_kcl9k_fll} reveals several noteworthy patterns.

First, the parameters found in Section~\ref{sec:supp_solver_config} allow all solvers to solve the exact same problem: all solvers agree within tolerance. The only number $> 1e-8$ here is in the ``olf'' row (Open Load Flow), for the worst-case mismatch.

Open Load Flow's (the default solver of pypowsybl) worst-case mismatch is $1.75e-06$ because it does not use the $\|\cdot\|_\infty$ norm internally but the $\|\cdot\|_2$ norm; this is why some values can occasionally exceed $1e-8$, even though the mean mismatch is always well below the tolerance ($4.4e-10 \ll 1e-8$).

PowerModels' convergence rate is relatively low compared to the others. This is why, in a real pipeline, the GENCO authors \citep{puech2026genco} preferred to use the more robust (but slower) mode of PowerModels to generate their data. We instead used the ``faster mode'' here, since our aim was not to generate data; had we needed to generate real power-flow data, we would have made the same choice as the GENCO authors and switched to the more robust Ipopt-based mode of PowerModels.

PGM (power grid models) does not converge for grids above 500 buses. It was primarily designed to model distribution grids, which have different properties than the transmission grids tested here; this likely explains why it diverges on the largest test cases. We intend to inform the PGM authors about this fact in the coming weeks.

The ``convergence rate'' of gpusim2grid is ``not applicable": as specified in Section~\ref{sec:suppl_gpusim2grid}, this model's stopping criterion is not based on accuracy (unlike the other baselines) but on the number of Newton-Raphson iterations (here varying from 0 to 7). As already detailed in the main paper, 5 iterations are enough in this setting to achieve an error lower than $1e-8$. Though gpusim2grid, like any physical solver, can detect internally the KCL mismatch (from its own representation of the system) and could raise "non-converging flags" if needed.

What is not reported in this table, however, is that sometimes, for some grids, some batch of gpusim2grid reports a ``NaN'' error. After a thorough investigation, we found that the sparse linear solver sometimes fails silently to find a good solution for the linear system and returns an unrealistically high value while reporting ``converged,'' causing subsequent iterations to overflow to floating-point ``NaN". This has already been reported to NVIDIA. More specific details, with examples of the misbehaving data, will be released alongside the paper's code for NVIDIA to have more data to work on if interested.

\begin{figure}
    \centering
    \includegraphics[width=0.5\linewidth]{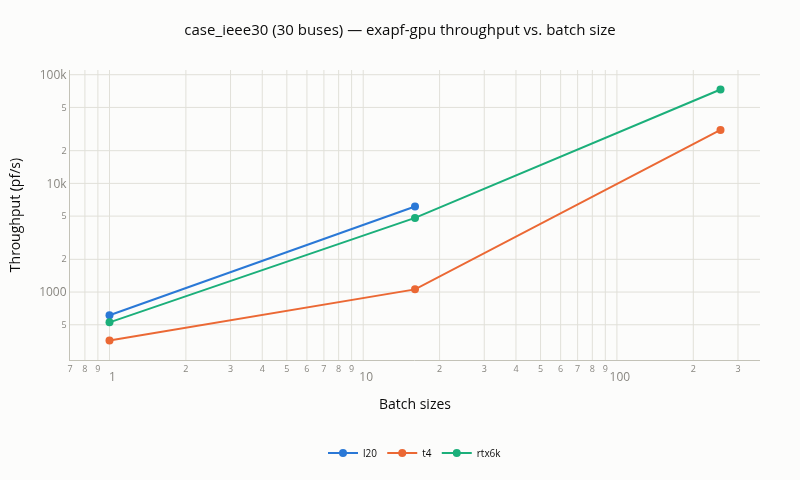}
    \caption{A specific run of exapf-gpu on case ieee\_30, showing similar issues we think are related to cuDSS (still in beta). The model is asked to solve $x$ powerflows with given input on 3 machines (each color). With batch sizes of 1 and 16, it succeeded on all machines and it manages to solve again the system when batch size is 256 on machine ``rtx6k'' and ``t4'' but not on the ``l20'' machine. Then all machines fail to solve the system with batch sizes of 512, 1024, etc. This is not due to an Out-of-Memory error, as exapf-gpu manages to successfully solve batches of 16,384 samples for case500, a much larger grid than this 30-bus case (figure not shown here, available on GitHub).}
    \label{fig:exapf_gpu_bug}
\end{figure}

We think that exapf, in GPU mode, suffers from the same type of issue. A more thorough analysis must be carried out, but Figure~\ref{fig:exapf_gpu_bug} is consistent with this explanation.

\begin{figure}[ht]
      \centering
      \includegraphics[width=0.62\linewidth]{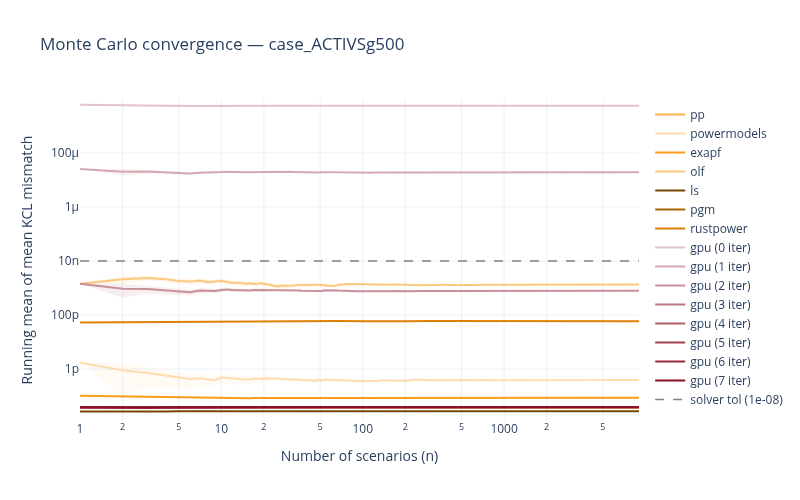}
      \caption{Representation, for case\_ACTIVSg500, of the ``mean mismatch'' for all baselines and all iterations of gpusim2grid. The x-axis reports the number of scenarios and the y-axis the running average (across all scenarios with id $\leq x$) of the error (plain lines), with its associated error bar (running standard deviation divided by $\sqrt{x}$, as is common practice in Monte-Carlo-like simulation). As shown here (full graphs will be available on GitHub), 9000 samples are large enough for the reported error to be representative of solver quality.}
      \label{fig:suppl_MC_KCL}
  \end{figure}

A final note on the ``arbitrary'' sample size of 9,000 chosen here: it was chosen so that the reported metric (mean mismatch) would be representative enough of the accuracy of the powerflows. We chose 9,000 and then checked with a Monte-Carlo-like estimate that the reported metric had indeed ``converged'' to its true value. Such graphs were produced for all 15 grids and both metrics; one example can be found in Figure~\ref{fig:suppl_MC_KCL}.

\subsection{Deep dive into the loads / generation scenarios}
\label{sec:suppl_cpu_spread}

The goal of this experiment was to benchmark the computation time of each solver when pushed to its limit. To that end, we decided to benchmark them on a rather simple task: one grid, many scenarios of demand and generation.

The process we wanted to benchmark is, for a given solver, a given grid, and a given scenario (loaded in main computer memory, RAM):
\begin{itemize}
    \item update the input of the solver with the new data
    \item solve the powerflow
\end{itemize}

As in \citet{puech2026genco} (Figure 7, page 15), we did not want to benchmark the time to load the data from disk or to initialize the solver into a usable state (\textit{e.g.}, loading its weights from the hard drive, compiling it, etc.).

This principle has some consequences:

\begin{itemize}
    \item In multi-threaded or multi-process experiments, the solver state often has to be initialized separately in each worker. Whenever possible, this initialization time is excluded from the measurements, as it would be negligible in operational settings where it is paid only once and then amortized over many simulations, for example when evaluating power flows under uncertainty in generation and demand.
    \item As an additional safeguard, we chose workloads so that each core spent about 30 seconds in the solve phase and processed a minimum number of power flows. This amortizes solver initialization or copy overhead over at least 30 seconds of actual computation, making its impact on the reported results negligible. As a result, the fastest solvers on the smallest grids required running several million power flows, likely far more than would arise in real grid operations.
    \item When dealing with a GPU model, we explicitly wanted our measurements to account for the time to transfer data between the GPU and main memory, in both directions. Indeed, in an operational process, data sitting still on the GPU has close to no use. It should be transferred back to the RAM for use by human operators or automatic processes alike.
    \item For JIT-compiled models (pandapower for Python, the Julia-based ones, or gpusim2grid) we first ran an untimed evaluation, then started the timing.
    \item We ``warmed up'' the GPU for all GPU-based solvers, for example making sure that all the shared libraries were loaded in RAM and that the model was ``transferred'' to the GPU, etc.
\end{itemize}

\subsubsection{Machine characteristics}
\label{sec:machine_charac}
We use three machines, each pairing a CPU and a GPU (4 GPUs of the stated model per machine): 

\begin{itemize}
    \item \textbf{l20} (2$\times$AMD EPYC 9755, 256 cores, $\sim$3\,TB RAM, NVIDIA L20 48GB VRAM)
    \item \textbf{t4} (2$\times$Intel Xeon Silver 4214R, 24 cores, $\sim$400\,GB RAM, NVIDIA Tesla T4 16GB VRAM)
    \item \textbf{rtx6k} (2$\times$AMD EPYC 9135, 32 cores, $\sim$800\,GB RAM, NVIDIA RTX PRO 6000 Blackwell 96GB VRAM).
\end{itemize}

All machines were on premise (not on the cloud), running RHEL 8.10 (Ootpa). Each machine uses ``hyperthreading'', so the available number of logical cores is double the number of physical cores (512 for the ``l20'' machine, 48 for the ``t4'' machine, and 64 for the ``rtx6k'' machine).

On ``t4'' and ``rtx6k", the Python version used was 3.12, and it was 3.13 on ``l20". The Julia version was 1.12.6 across all machines.

The (main) packages used were the same across all machines:

\begin{itemize}
    \item pandapower: 3.5.4
    \item powermodels: 0.21.6 (julia)
    \item exapf: 0.13.0 (julia)
    \item pypowsybl / olf: 1.16.1
    \item lightsim2grid: 1.0.0
    \item pgm: 1.13.141
    \item rustpower: 0.5.0
    \item pytorch: 2.12.1+cu130
    \item torch geometric: 2.8.0.post1
\end{itemize}

In principle, all experiments were run on all 3 machines. We only report results for the ``rtx6k'' machine because we think it's the ``most standard'' machine of the 3. ``t4'' is rather old and approaching end of life, the ``l20'' has 512 available CPU threads (2 times 256 real physical cores) which is not the most common hardware.

We still wanted to make as many runs on the 3 different machines as possible to ensure that the results we have are robust to different hardware specification.

\subsubsection{Highest throughput for CPU solvers}

\begin{figure}
    \centering
    \includegraphics[width=0.5\linewidth]{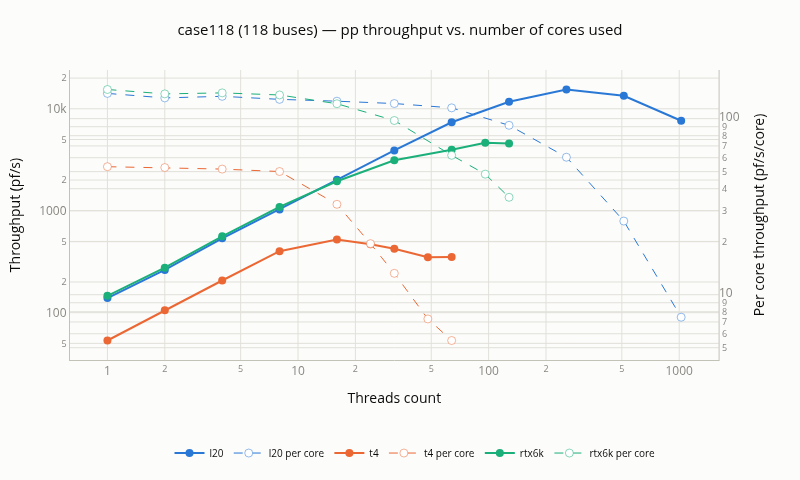}
    \caption{Example of one of the $15 \text{ (grids)} \times 7 \text{ (solvers)} = 105$ plots generated, where a given method (in this case pandapower) is timed on one grid (in this case case118) on 3 machines: ``l20", ``rtx6k'' and ``t4". The x-axis represents the number of cores used and the y-axis (left) represents the throughput (number of powerflows per second). The maximum throughput for this method on this grid is ~15200 for the ``l20'' machine, ~4700 for the ``rtx6k'' machine, and ~550 for the ``t4'' machine.}
    \label{fig:pp_timing_example}
\end{figure}

To obtain the highest throughput for the CPU solvers, we ran all 7 solvers using different core counts (from 1 up to 1024 for some experiments) for each grid. If the solver provides a native parallelism feature, we used it; otherwise, we fell back to the native ``parallel'' tool available in the language (\textit{i.e.}, multiprocessing or multithreading for Python, or native threads in Julia). We ran most of our experiments on 3 machines at our disposal to generate figures similar to Figure~\ref{fig:pp_timing_example}. In parallel experiments, we disable internal threading in the benchmarked solvers to avoid thread oversubscription. This matters when a method relies on libraries such as OpenMP, BLAS, or MKL for linear algebra. As a safeguard, we set \texttt{OMP\_NUM\_THREADS}, \texttt{OPENBLAS\_NUM\_THREADS}, \texttt{MKL\_NUM\_THREADS}, \texttt{NUMEXPR\_NUM\_THREADS}, \texttt{VECLIB\_MAXIMUM\_THREADS}, and \texttt{RAYON\_NUM\_THREADS} to 1.

\begin{figure}
    \centering
    \includegraphics[width=0.5\linewidth]{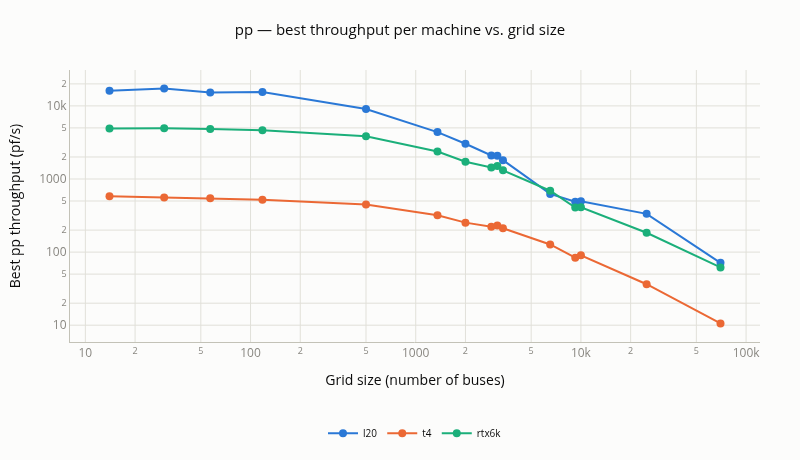}
    \caption{Example of one of the $7$ plots generated from the aggregation of Figure~\ref{fig:pp_timing_example} across all grids. The x-axis is the grid size (15 different grids, from 14 to 70000 buses) and the y-axis is the maximum throughput, maximum taken over the different number of threads/processes simulated (see Figure~\ref{fig:pp_timing_example}).}
    \label{fig:pp_timing_grids_example}
\end{figure}

Each of these figures is then summarized per grid size to give the maximum throughput per method across all grids, yielding a figure similar to Figure~\ref{fig:pp_timing_grids_example}.\\

\begin{figure}
    \centering
    \includegraphics[width=0.5\linewidth]{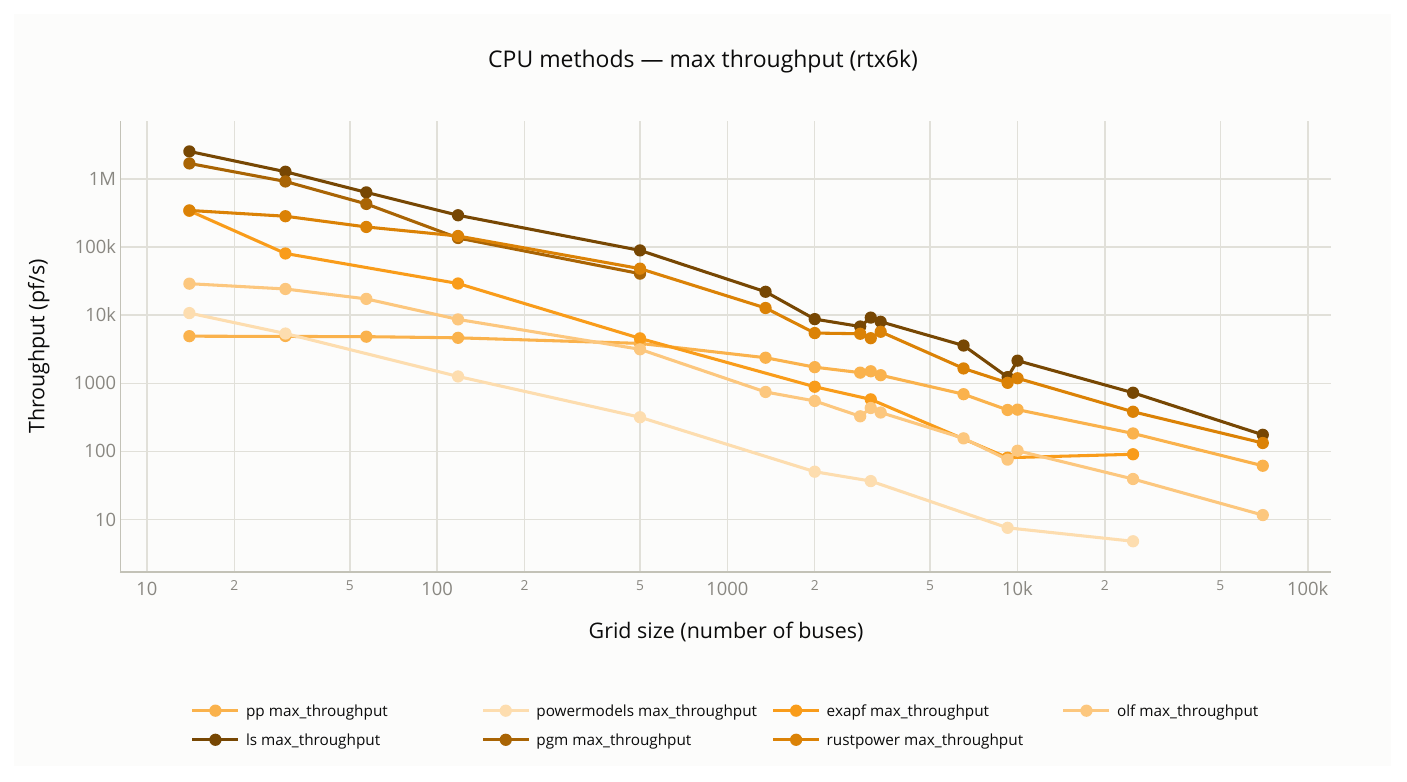}
    \caption{Maximum throughput obtained for the 7 CPU solvers on the rtx6k machine. As we can see, some solvers are more than 100x faster than others for this given task.}
    \label{fig:20260826_cpu_max_rtx6k}
\end{figure}

We now have 3 different plots, one for each machine, presenting the maximum throughput for each of the 7 CPU solvers benchmarked; results are available in Figure~\ref{fig:20260826_cpu_max_rtx6k}. \\

We then report, in the main Figure~\ref{fig:exp_best}, the lightsim2grid curve, since it is the one with the highest measured throughput for the selected ``rtx6k'' machine. \\

Making this figure requires approximately 4700 CPU-hours ($\approx$195 core-days) for the final runs. Most of the time was spent on powermodels (the slowest of the baselines in this setting). 630M power flows have been run for this experiment, most of them by pgm and lightsim2grid on smallest grids: the fastest solver in this benchmark.

\paragraph{Caveats}

We want to emphasize some limits of this benchmark methodology. We did not try to optimize each solver for speed so as to achieve its absolute best performance, which would have been mandatory if we wanted to rank solvers based on their throughput, which is \textbf{not} our goal here. Our approach was centered around a ``best effort'' strategy: solvers were optimized for speed using easily available/documented features, as we believe is common practice in the ``machine learning for power systems'' community.

Some solvers (pgm, lightsim2grid) natively offer features matching almost 100\% of our benchmark here (scenarios where loads and generation change but not the topology). We decided to use this capability for these two solvers. Our benchmark does not try to rank these CPU solvers, but rather to show what is possible with a CPU; this is why we used the native feature whenever available.

Conversely, some other solvers (\textit{e.g.}, pandapower, powermodels, pypowsybl/OLF) ``start from scratch'' each time a new scenario is computed, for example by rebuilding the entire $Y$ matrix and $s$ vector. This might be rather inefficient, but this absolutely does not mean these solvers are slow in general; it only means they have not been optimized for the specific task used in this benchmark. They could shine on other types of tasks not benchmarked here, for example contingency analysis.

We were more familiar with Python than with Julia. Following our ``best effort'' principle, we did not try to make our Python code overly complex, though it might naturally be more efficient simply because of our familiarity with the language. This would have been an issue had we tried to rank CPU solvers, which has never been the goal of this benchmark.

\subsubsection{Highest throughput for GPU / ML solvers}

We adopted a similar methodology to find the best throughput of the methods using GPUs (Newton-Raphson-based solvers and machine-learning surrogates), but instead of varying the number of CPU cores, we varied the batch size.

For learned models, we have not tried to retrain them. We only used published architectures and their hyperparameters (number of ``layers'', size of each ``layer'', etc.). We are only concerned with evaluation throughput and this is why we also deactivated the computation of gradients for these models.

Usually, we generated a dataset consisting of 16,384 rows (for each grid size). Then, on this dataset, we ran the same model for batch sizes 1, 2, 4, ..., 16384. To save computation time, we added two break conditions: a) if there are more than 100 batches, only the first 100 batches are simulated and the average is reported over these 100 batches; b) if a model reports an ``Out of Memory'' error on the device for a given batch size, the larger batches are not simulated.

By diving into the computation time, we also found some optimizations that can be made to PF-$\Delta$'s useful \texttt{InMemoryDataset}, which we will share with the community, notably regarding faster preprocessing in the coming weeks.

\begin{figure}
    \centering
    \includegraphics[width=0.5\linewidth]{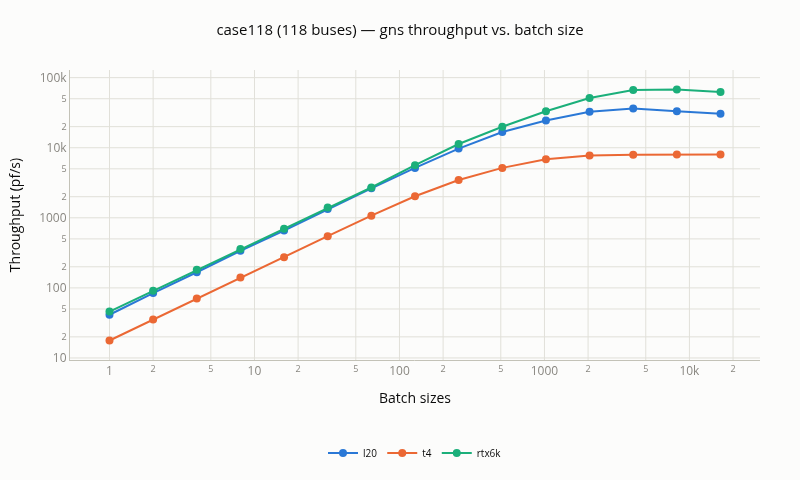}
    \caption{Example of one of the $15 \text{ (grids)} \times 5 \text{ (solvers)} = 75$ plots generated, where a given method (in this case Graph Neural Solver) is timed on one grid (in this case case118) on 3 machines: ``l20", ``rtx6k'' and ``t4". The x-axis represents the batch size used and the y-axis represents the throughput (number of powerflows per second). The maximum throughput for this method on this grid is 36000 for the ``l20'' GPU, 65000 for the ``rtx6k'' GPU, and 7900 for the ``t4'' GPU.}
    \label{fig:gns_timing_example}
\end{figure}

First, we produced figures for a given method on a given grid, as shown in Figure~\ref{fig:gns_timing_example}.

Then we aggregate the results for a given method across all computed grids:

\begin{figure}
    \centering
    \includegraphics[width=0.5\linewidth]{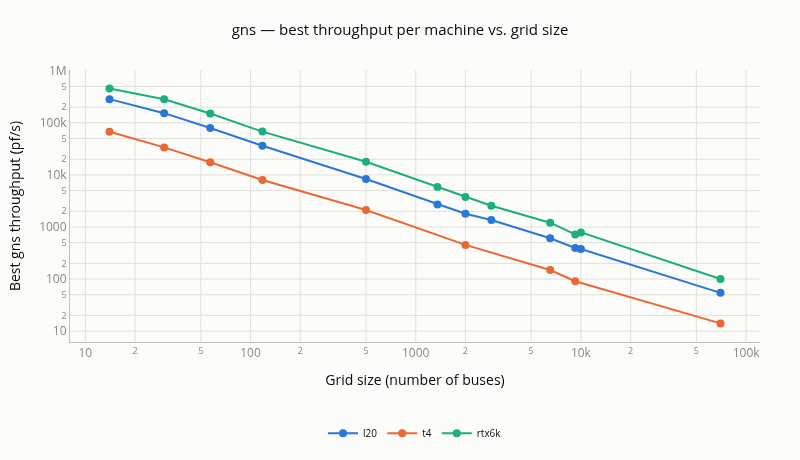}
    \caption{Example of one of the $5$ plots generated from the aggregation of Figure~\ref{fig:gns_timing_example} across all grids. The x-axis is the grid size (15 different grids, from 14 to 70000 buses) and the y-axis is the maximum throughput, maximum taken over the different batch sizes tested (see Figure~\ref{fig:gns_timing_example}).}
    \label{fig:gns_timing_grids_example}
\end{figure}

\begin{figure}
    \centering
    \includegraphics[width=0.5\linewidth]{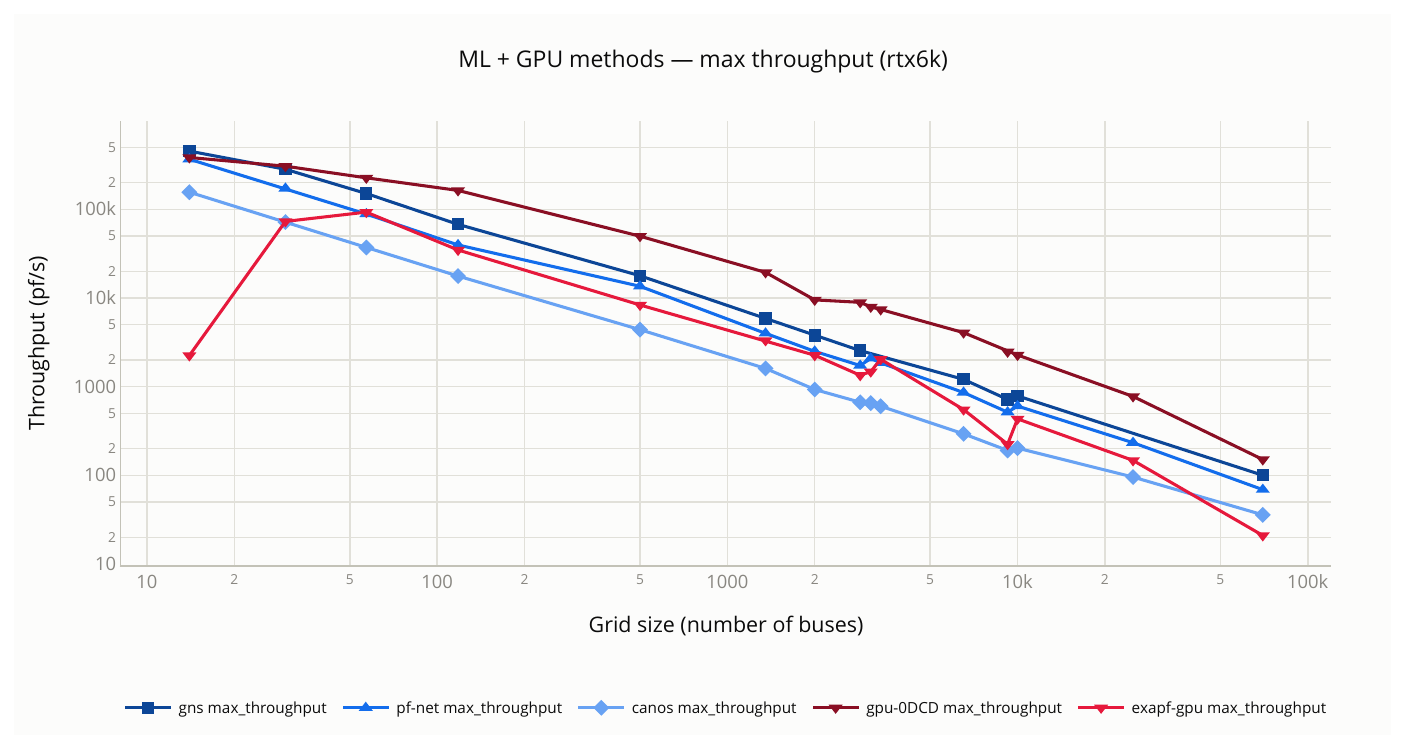}
    \caption{Maximum throughput obtained for the 5 GPU solvers (2 Newton-Raphson and 3 machine-learning) on the rtx6k machine.}
    \label{fig:20260826_gpu_rtx6k}
\end{figure}

We now have 3 different plots, one for each machine, presenting the maximum throughput for each of the 5 GPU solvers benchmarked (2 Newton-Raphson and 3 machine-learning); results are available in Figure~\ref{fig:20260826_gpu_rtx6k}.

We then pick the best performing method in each category from this plot to include it in the Figure \ref{fig:exp_best} of the main paper: Graph Neural Solver is the fastest learned model and gpusim2grid is the fastest GPU solver available.

In Figure~\ref{fig:20260826_gpu_rtx6k} we can see the phenomenon described in Section~\ref{sec:supp_kcl_baselines} regarding exapf (and most likely cuDSS instability). As shown in Figure~\ref{fig:exapf_gpu_bug}, for some grids (like case14, case\_ieee30, and case57) cuDSS fails to solve the systems. This limits the batch size to only 256 (larger sizes diverge for these grids, for reasons that remain unclear), which is not enough to saturate the GPU on these small grids, leading to poor throughput for exapf on the 3 smallest grids.

We can also notice a small artifact of our benchmark: for the smallest cases, too few scenarios were simulated for gpusim2grid to saturate the GPU. As a result, the curve is flat at the beginning of the plot, whereas it should decrease linearly on this log-log scale.

\end{document}

%% file: domain_checklist.tex

\section*{AI4PowerGrids Domain Checklist}

\begin{enumerate}
  \item \textbf{Physics feasibility.} All results reported for every solver use the full AC power-flow equations (Eq.~\ref{eq:powerflow}), not a DC or linearized approximation. Feasibility is assessed with an independent reference KCL evaluator (Section~\ref{sec:formulation}, detailed in Section~\ref{sec:suppl_kcl_checker}), reporting mean and worst-case constraint-violation statistics rather than aggregate error alone (Table~\ref{tab:kcl_9k}, full per-solver results in Table~\ref{tab:suppl_kcl9k_fll}).

  \item \textbf{Out-of-distribution evaluation.} No solver benchmarked here is trained by us, and the physics-based solvers require no training at all, so ``out-of-distribution'' in the usual train/test sense does not directly apply to them. That said, the benchmark itself is a structural distribution shift: all methods, including the published ML surrogates, are evaluated across 15 grids spanning 3.5 orders of magnitude in size (14 to 70{,}000 buses, Section~\ref{sec:common_settings}), and, for the PF-$\Delta$ comparison, across four topology/operating regimes -- normal ($N$), single- and double-contingency ($N$-1, $N$-2), and close-to-infeasible (C2I) -- none of which match the ML surrogates' own training conditions (Section~\ref{sec:pfdelta_accuracy}, Table~\ref{tab:pfdelta-accuracy-mean}). We report this as the closest available proxy for OOD evaluation given that no method here was trained by us.

  \item \textbf{Failure modes.} Convergence rate per solver is reported in Table~\ref{tab:kcl_9k}, with the complete per-solver, per-grid breakdown and discussion of failure modes in Table~\ref{tab:suppl_kcl9k_fll} (Section~\ref{sec:supp_kcl_baselines}), including power grid models' failure to scale beyond small grids and the GPU NaN-mismatch cases tied to a cuDSS beta issue. Diverging cases are explicitly retained (not filtered out) in Table~\ref{tab:pfdelta-accuracy-mean} and discussed in the accompanying results paragraph (Section~\ref{sec:pfdelta_accuracy}).

  \item \textbf{System-level effect.} Not applicable. This paper benchmarks standalone AC power-flow solving -- a computational building block used by every grid operator -- and makes no claim about downstream effects when PF is embedded in a larger control loop (e.g., contingency screening, OPF, or real-time dispatch). We state this explicitly as a scope limitation in Section~\ref{sec:discussion} rather than leaving it unaddressed.

  \item \textbf{Data and code.} Code for gpu-0DCD is already available at an anonymized repository (\url{https://anonymous.4open.science/r/gpusimulator-0DCD}), which will also host the independent KCL evaluator and the full benchmark harness (physics-based and ML solvers) by the time of review. We do not plan to release the sampled scenario data itself: the sampling methodology is simple enough to reproduce locally, and regeneration is cheaper than storage or transfer. As a reference point, extrapolating storage costs from the GridFM case14 dataset (\url{https://huggingface.co/datasets/gridfm/pf_small_case14_ieee/tree/main}, 400MB for 200{,}000 instances) to the 8{,}928{,}571 case14 scenarios solved for this paper would require roughly 178GB; regenerating that same data on demand takes about 1 minute on a single core and under 3 seconds using multiple cores with the fastest solver benchmarked here.
\end{enumerate}